\documentclass[prl,twocolumn,epsfig,rotate,showpacs]{revtex4}
\usepackage{epsfig}
\usepackage{amsmath}
\usepackage{graphicx}
\usepackage{dcolumn}
\usepackage{bm}
\begin{document}

\title{Energy transfer process in gas models of Lennard-Jones interactions}
\author{Jinghua Yang }
\author{Yong Zhang}
\author{Jiao Wang}
\author{Hong Zhao}
\email{zhaoh@xmu.edu.cn}
\affiliation{Department of Physics and Institute of Theoretical Physics and Astrophysics,\\
Xiamen University, Xiamen 361005, China. }
\date{\today}

\begin{abstract}
We perform simulations to investigate how the energy carried by a molecule transfers to others in an equilibrium gas model. For this purpose we consider a microcanonical ensemble of equilibrium gas systems, each of them contains a tagged molecule located at the same position initially. The ensuing transfer process of the energy initially carried by the tagged molecule is then exposed in terms of the ensemble-averaged energy density distribution. In both a 2D and a 3D gas model with Lennard-Jones interactions at room temperature, it is found that the energy carried by a molecule propagates in the gas ballistically, in clear contrast with the Gaussian diffusion widely assumed in previous studies. A possible scheme of experimental study of this issue is also proposed.
\end{abstract}

\pacs{05.60.Cd, 51.10.+y, 51.20.+d}
\maketitle

One important task of statistical mechanics is to understand various
transport processes. A well-known successful example is the
self-diffusion of gas molecules: Due to Einstein's 1905 work, it has
been widely accepted that a particle in a gas undergoes the Brownian
motion \cite{Einstein}, which can be modeled essentially with the
random walk \cite{smoluchowski} and the resulting probability
distribution function (PDF) follows the diffusion equation
\cite{Einstein,smoluchowski}. Because of its fundamental importance
for various scientific disciplines, the study of Einstein's theory
has never ceased. Very recently, the direct experimental
measurement of the instantaneous velocity of a Brownian particle in
a gas has been realized, and the random walk picture was confirmed with high precision \cite{scieceexpress}.

Another question of fundamental importance is how the energy carried
by a molecule transfers with time, which is a key step towards understanding the macroscopic energy transport. In general, the existing theories, taking for example the Helfand theory \cite{helfand}, approached this issue by simply extending the random walk picture of a Brownian particle, and predicted a Gaussian energy density distribution as well. However, it should be pointed out that whether random walk is the underlying mechanism of the energy dispersion has never been examined experimentally nor numerically in a gas or more generally in fluids.

Unlike in the study of the self-diffusion (or mass diffusion) where the position of a particle can be traced accurately \cite{expt-1,expt-2,expt-3,expt-4} (and now even its instantaneous velocity can be measured \cite{scieceexpress}), a key difficulty in the study of the energy transfer is that it is hard to trace the energy transferred from particle to particle. The mass diffusion can be explored by focusing on the trace of an individual particle, but by nature, the energy transfer is a collective behavior involving all the molecules related by the transferred energy.

In this work we perform an equilibrium molecular dynamics investigation to explore how the energy of a molecule may transfer in a gas. We will restrict ourselves to a 2D gas model, but it has been verified that in its 3D counterpart the results remain qualitatively the same. We assume that the gas is composed of only one kind of molecule with diameter $\sigma$ and mass $m$.  The setup consists of a square of area $S$ with periodic boundary conditions and $N$ molecules moving inside. The interaction between molecules is given by the Lennard-Jones potential and the Hamiltonian of the system reads
\begin{equation}  \label{eq:one}
H=\sum_{i}^{N}H_{i}=\sum_{i}^{N}\{\frac{\mathbf{p}_{i}^2}{2m}+\sum_{j=1(\neq
i)}^{N}{2\varepsilon [(\frac{\sigma}{r_{ij}})^{12}-(\frac{\sigma}{r_{ij}}%
)^{6}]}\},
\end{equation}
where $\varepsilon$ is a constant governing the interaction strength and $r_{ij}$ denotes the distance between molecules $i$ and $j$. In our calculations the dimensionless parameters are set to be $\sigma=1$, $m=1$, $\varepsilon=1$ and the Boltzmann constant $k_{B}=1$; In addition, the number of the molecules $N=2500$ and the area $S=200\times 200$ are adopted. Another important parameter is the temperature, which is fixed at $T=2.5$, a value that corresponds to the room temperature with other adopted parameter values. To make the simulations more efficient, the potential energy between two molecules is approximated by zero when their distance is larger than $r_{c}=3.5$, as conventionally adopted in the molecular dynamics studies of gases.
Given these, the evolution of a system can be simulated straightforwardly. In our calculations the 7-order Runge-Kutta algorithm \cite{runge} with step 0.01 is employed.

\begin{figure*}[tbp]
\epsfig{file=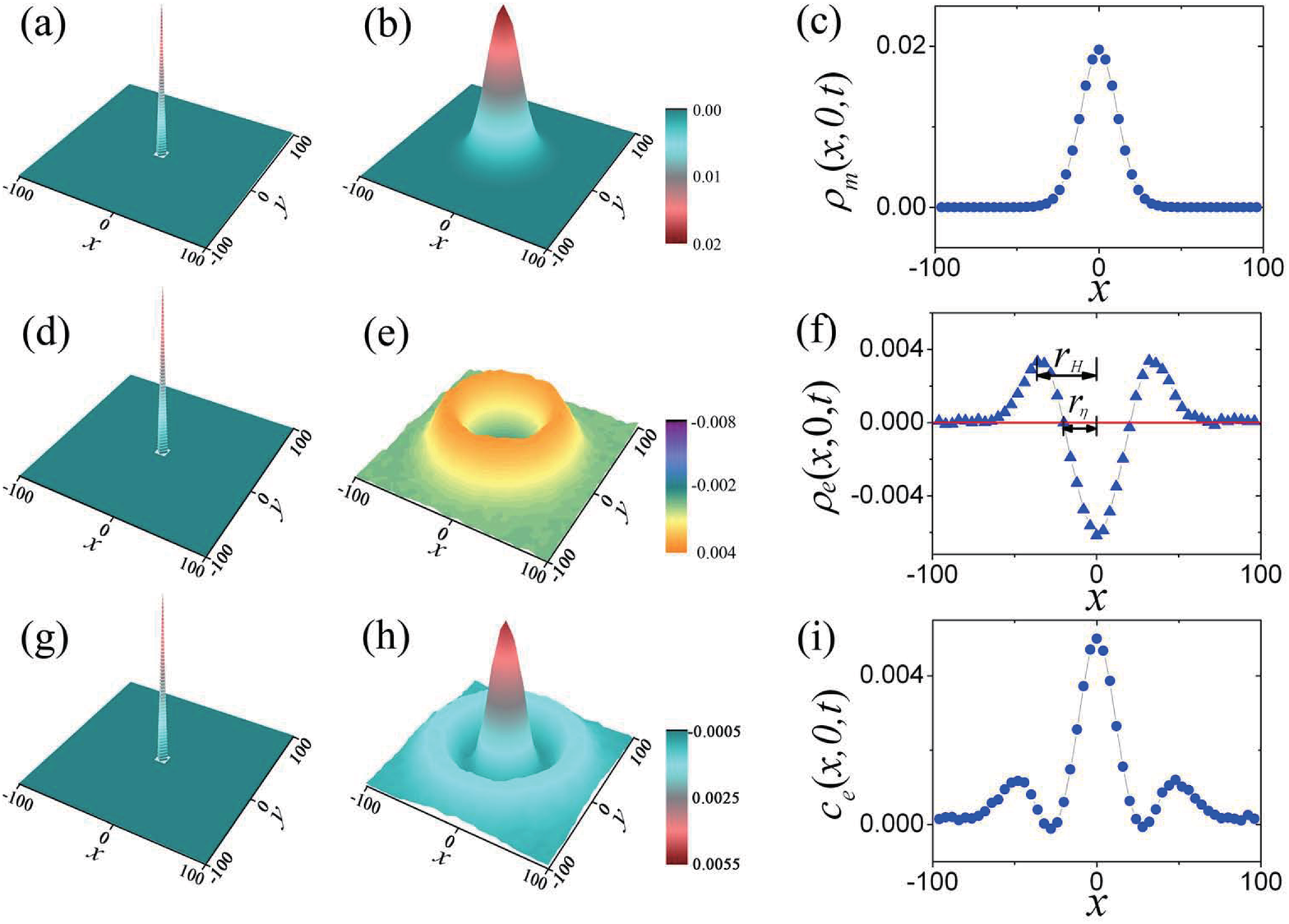,width=12cm}
\caption{The PDF $\protect\rho_{m}(\mathbf{r},t)$ of the tagged molecule at time $t=0$ (a) and $t=15$ (b). (For the sake of presentation, the coarse-grained results over a grid of squares of size $4\times 4$ are plotted.) The intersection of the plot in (b) with plane $y=0$ is shown in (c) for a close look, where $\protect\rho_{m}(x,y,t)\equiv \protect\rho_{m}(\mathbf{r},t)$. (d)-(f) and (g)-(i) are the same as (a)-(c) but for the ensemble averaged energy density distribution $\protect\rho_{e}(\mathbf{r},t)$ and the spatiotemporal correlation function $c_{e}(\mathbf{r},t)$ of the energy fluctuation respectively. The ensemble average for all the three cases is evaluated over $3\times10^{8}$ systems.}
\label{fig1}
\end{figure*}

The ensemble is prepared through the following
three steps: (i) First, a ``seed'' equilibrium system of $N$ molecules
at temperature $T$ is prepared by evolving a system for a long
enough time ($>1\times 10^{6}$) from a properly assigned random
state; (ii) Then a molecule is chosen and its position is set to be the origin by translating the coordinate system. In this way we build an equilibrium system with one molecule initially localized at the origin. The molecule at the origin is hereinafter referred to as
``the tagged molecule''. By assigning respectively all $N$ molecules to be the tagged one in this way, i.e. setting their initial positions to be at the origin one by one, a subensemble of $N$ equilibrium systems, each has a tagged molecule initially residing on the origin, is thus prepared. (iii) By repeating step (i) to have different realizations of the seed equilibrium system followed by (ii) to generate the corresponding subensemble, we then build the whole ensemble whose member number can be large enough (up to $\sim 10^8$) for satisfying statistic results. It is worth noting that building the ensemble in such a way is not new; a similar idea was once employed by Helfand to establish the energy diffusion theory of fluids \cite{helfand}.

To investigate that as the system evolves, how a molecule (represented by the tagged molecule) diffuses and how the energy it carries initially spreads over the whole system, we will study in particular the following three processes with the ensemble prepared:

A. The self-diffusion, or mass diffusion process, which can be accessed by calculating the PDF, denoted by $\rho_{m}(\mathbf{r},t)$, of the tagged molecule. It is defined as
\begin{equation}  \label{rhom}
\rho_{m}(\mathbf{r},t)\equiv \langle\delta[\mathbf{r}-{\mathbf{r}}_{1}(t)] \rangle,
\end{equation}
where $\langle\cdot\rangle$ denotes the ensemble average, ${\bf{r}}_{i}(t)$ denotes the position of molecule $i$ at time $t$ and number one molecule represents the tagged molecule. As initially the tagged molecule is located at the origin, i.e., ${\bf{r}}_{1}(0)=\mathbf{0}$, we have $\rho_{m}(\mathbf{r},0)=\delta(\bf{r})$.

B. The dispersion of the energy initially carried by the tagged molecule. For this aim we consider the ensemble averaged energy density distribution (EAEDD) of the systems, i.e.
\begin{equation}
\langle E(\mathbf{r},t)\rangle =\langle H_{1}(t)\delta \lbrack \mathbf{r}-%
\mathbf{r}_{1}(t)]\rangle +\langle \sum_{j=2}^{N}H_{j}(t)\delta \lbrack
\mathbf{r}-\mathbf{r}_{j}(t)]\rangle .  \label{eq:two}
\end{equation}%
Initially, as the origin is occupied exclusively by the tagged molecule, we have $\langle H_{1}(0)\delta [{\bf{r}}-{\bf{r}}_{1}(0)]\rangle =\widetilde{E}\delta (\mathbf{r})$,
where $\widetilde{E}\equiv\langle H_{j}\rangle$ is the average energy of a molecule in the gas. On the other hand, as the rest area of the space ($\mathbf{r}\neq \mathbf{0}$) is occupied uniformly by other $N-1$ molecules, the EAEDD they contribute to, i.e. the second term on the r.h.s of Eq. (\ref{eq:two}), equals a constant $\eta \equiv \widetilde{E}(N-1)/S$. Hence initially $\langle E(\mathbf{r},0)\rangle $ is characterized by a peak at the origin with a flat background. As the system evolves, while the portion of $\langle E(\mathbf{r},t)\rangle $ contributed by the tagged molecule may spread out from its initial $\delta $-function, that by the other $N-1$ molecules remains to be $\eta $. Therefore we can use the reformed distribution $\rho _{e}(\mathbf{r},t)\equiv (\langle E(\mathbf{r},t)\rangle -\eta )/\widetilde{E}$ to capture the  dispersion of the energy initially carried by the tagged molecule. This is the key technique of this work; with it the energy transferred to others from a $single$ molecule can thus be traced by checking the EAEDD of the $whole$ system, making it possible to study the former conveniently.

C. The energy fluctuation correlation. The spatiotemporal correlation function of the energy fluctuation \cite{zhao} is a useful tool \cite{Lepri,Dhar} in tracing how the energy transfers \cite{notelead,lead}. When the energy initially concentrated at the origin (the cause) is transferred to position $\mathbf{r}$ at time $t$, it will induce a solid correlation (the effect); we have to expose the correlation induced exclusively by this causality. As the gas model we study here is a microcanonical system with the total energy conserved, we have at any time $\tau$ that $\sum_{j=1}^{N}\Delta H_{j}(\tau)=0$ and thus $\Delta H_{1}(\tau)\sum_{j=1}^{N}\Delta H_{j}(\tau)=\Delta H_{1}(\tau)\Delta H_{1}(\tau)+\sum_{j=2}^{N}\Delta H_{1}(\tau)\Delta H_{j}(\tau)=0$. Here $\Delta H_{j}(\tau)\equiv H_{j}(\tau)-\widetilde{E}$. Considering the  ensemble average, we then have $\langle \Delta H_{1}(\tau)\Delta H_{j}(\tau)\rangle =\langle \Delta H_{1}^{2} (\tau)\rangle/(N-1)$
since the gas is homogeneous, which indicates that there is a trivial correlation between any two molecules induced by the conservation of the energy. To get rid of it we consider instead $\langle \Delta H_{1}(0)\sum_{j=1}^{N}\Delta H_{j}(t)\delta \lbrack \mathbf{r}-\mathbf{r}_{j}(t)]\rangle$;
As initially (at $t=0$) it has a center of $\delta $-function form $\langle \Delta H_{1}(0)\Delta H_{1}(0)\delta (\mathbf{r)}\rangle $ and a flat background $\mu\equiv -\langle \Delta H_{1}(0)\Delta H_{1}(0)\rangle N/(N-1)S$, we accordingly define the spatiotemporal correlation function as $c_{e}(\mathbf{r},t)\equiv \langle \Delta H_{1}(0)\sum_{j=1}^{N}\Delta H_{j}(t)\delta \lbrack \mathbf{r}-\mathbf{r}_{j}(t)]\rangle -\mu $ to explore the correlation at position $\mathbf{r}$ and time $t$ induced by the initial energy fluctuation $\Delta H_{1}(0)$.

The main results are summarized in Fig. 1. First of all Fig. 1 (a)-(c) are for $\rho_{m}(\mathbf{r},t)$, the PDF of the tagged molecule. Initially it is a $\delta$-function as seen in Fig. 1(a), in agreement with the fact that the tagged molecule is located at the origin at the beginning. Later it develops into a Gaussian distribution [Fig. 1(b)-(c)]. As a double check we have also studied the squared displacement of the tagged molecule and found that it depends on time linearly after a transient time of $t\approx 10$; i.e., $\langle|\mathbf{r} |^{2}(t)\rangle=4Dt$ with $D=7.08\pm 0.01$ as suggested by the best linear fitting $\langle|\mathbf{r} |^{2}(t)\rangle$ over $10<t<80$. These results are clear evidence that the self-diffusion of a molecule is normal in our system.

Fig. 1 (d)-(f) show the energy transfer behavior. The $\delta$-function seen in Fig. 1 (d) indicates that the energy we are interested in is initially located at the origin. However, in its later development, a distinctive difference from the self-diffusion of a molecule can be identified: Instead of a Gaussian distribution, $\rho_{e}(\mathbf{r},t)$ features a growing ``crater'', i.e., a ring ridge (where $\rho_{e}(\mathbf{r},t)>0$) moves outwards leaving behind a dip (where $\rho_{e}(\mathbf{r},t)<0$) in center [see Fig. 1(f)].

Fig. 1 (g)-(i) show the correlation function
$c_{e}(\mathbf{{r}},t)$. It can be found that
$c_{e}(\mathbf{{r}},t)$ reveals the same features of the energy
transfer process seen in $\rho
_{e}(\mathbf{{r}},t)$. The crater structure in
$c_{e}(\mathbf{{r}},t\mathbf{)}$ appears at the same position and
expands outward with the same velocity as that in $\rho
_{e}(\mathbf{{r}},t)$. As in the center dip region of $\rho
_{e}(\mathbf{{r},}t)$ we have $\langle \Delta H_{j}(t)\rangle <0$ [see Fig. 1(f)], which implies $\langle \Delta H_{1}(0)\Delta H_{j}(t)\rangle >0$, the center peak in $c_{e}(\mathbf{{r},}t)$ is  consistent with the dip in $\rho _{e}(\mathbf{{r},}t)$. It is interesting to note that
between the crater and the center peak there is a small region
showing a negative correlation; This feature is different from that
observed in the lattice models, where
$c_{e}(\mathbf{{r}},t\mathbf{)}$ is always positive and as a consequence can be employed to represent the PDF of the energy diffusion \cite{zhao}.

\begin{figure}[tbp]
\vspace{-0.1cm} \epsfig{file=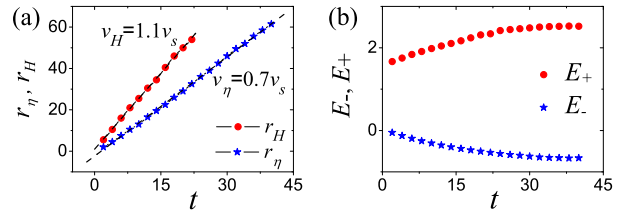,width=9cm}\vspace{-0.4cm}
\caption{(a) The time dependence of the radii characterizing the crater structure in $\rho _{e}({\bf r},t)$, corresponding to the top ring of the ridge (red bullets) and the opening of the center dip (blue stars) respectively. The best fittings (dashed lines) suggest their expanding speeds $\protect\nu_{H}\approx1.1\protect\nu_{s}$ and $\protect\nu%
_{\protect\eta}\approx0.7\protect\nu_{s}$.
(b) The time dependence of the positive and negative potion of energy, i.e. $E_{+}$ (red bullets) and $E_{-}$(blue stars),
corresponding to the integrated energy density distribution over region $|\mathbf{r}|>r_{\protect\eta}$ and $|\mathbf{r}|<r_{\protect\eta}$,
respectively. }
\label{fig2}
\end{figure}

Now let us have a closer look at $\rho_{e}(\mathbf{r},t)$, the main result of this paper. Two key geometric parameters characterizing the crater structure of $\rho _{e}(\mathbf{r},t)$ as shown in Fig. 1(e)-(f), denoted by $r_{\eta }$ and $r_{H}$ respectively, are the radius of the intersection ring on which $\rho _{e}(\mathbf{r},t)=0$ and that of the ring of the ridge top where $\rho _{e}(\mathbf{r},t)$ takes the maximum value [see Fig. 1(f)]. It is found that they both depend on time linearly [see Fig. 2(a)], but however, the speed of the ridge top, represented by $\nu _{H}\equiv dr_{H}/dt$, is different from that of the opening of the dip $\nu _{\eta}\equiv dr_{\eta }/dt$: The best linear fitting results suggest $\nu _{H} \approx 1.6 \nu _{\eta}$. Just as a comparison, it is interesting to note that the two speeds are comparable to the speed of sound, a macroscopic characteristic; i.e., $\nu _{H}\approx 1.1\nu _{s}$ and $\nu _{\eta }\approx 0.7\nu _{s}$, where $\nu _{s}=\sqrt{c_{p}k_{B}T}/\sqrt{c_{v}m}$ is the sound speed   of the 2D ideal gas of the same molecular mass and density and at the same temperature as our model. On the other hand, as $\langle E(\mathbf{r},t)\rangle -\eta $ describes how the initial energy carried by the tagged molecule transfers away, the interesting fact that $ \rho _{e}(\mathbf{r},t)$ has a negative center suggests that during this process some energy of the neighboring molecules is brought away in addition. This additional portion of energy is given by $-E_{-}$ where $
E_{-}\equiv \widetilde{E}\int_{|\mathbf{r}|<r_{\eta }}\rho _{e}(\mathbf{r},t)d\mathbf{r}$. Similarly, the total positive energy carried by the bulk of the ridge is given by $E_{+}\equiv \widetilde{E}\int_{|\mathbf{r}|>r_{\eta}}\rho _{e}(\mathbf{r},t)d\mathbf{r}$. Due to the conversation of the
energy we have always $E_{+}+E_{-}=\widetilde{E}$. Fig. 2 (b) shows the time dependence of $E_{-}$ and that of $E_{+}$; initially $E_{-}$($E_{+}$) decreases (increases) but after a transition time it reaches a constant. In other words, eventually the total energy brought away by the ridge is a constant and is larger than the energy initially the tagged molecule carries. Together with the results of $\nu _{H}$ and $\nu _{\eta }$, they suggest clearly that rather than the Gaussian diffusion, the energy carried by a molecule \textit{propagates} away ballistically in our gas model.

Why the energy transfer and the molecule self-diffusion are so different can be actually understood within the framework of the random walk theory. According to this theory, a normal diffusion occurs if the random walker loses its memory of the previous states completely, otherwise an abnormal diffusion may take place instead. In our gas system, if we focus on the tagged molecule, we may find that due to its frequent collisions with others, its memory of the initial direction of motion suffers a quick loss. This memory loss process can be measured by the decay of the autocorrelation function $A(t)\equiv \langle \mathbf{p}_{1}(0)\cdot \mathbf{p}_{1}(t)\rangle $ of the tagged molecule. (Here $\mathbf{p}_{1}(t)$ denotes the momentum of the tagged molecule.) Indeed, as shown in Fig. 3(a), $A(t)$ decreases exponentially with time, hence the motion of the tagged molecule is equivalent to that of a random walker.

However, the information of the initial moving direction of the tagged molecule is well remembered by the $whole$ system. When the energy of the tagged molecule transfers to others, the memory of its initial state may transfer to the surrounding molecules as well. To show this we consider the correlation function $C(t)\equiv \sum_{j=1}^{N}\langle \mathbf{p}_{1}(0)\cdot \mathbf{p}_{j}(t)\rangle $ as a measure of the total memory (note that $C(t)$ is evaluated over the whole system). Dividing the momentum of a molecule, say molecule $j$, into
two parts; i.e., $\mathbf{p}_{j}(t)=\mathbf{p}_{j}^{\prime }(t)+\mathbf{p}_{j}^{\prime \prime }(t)$ , where $\mathbf{p}_{j}^{\prime }(t)$ and $\mathbf{p}_{j}^{\prime \prime }(t)$ represent respectively the momentum transferred to molecule $j$ from the tagged molecule and other molecules, we then have $C(t)=\sum_{j=1}^{N}\langle \mathbf{p}_{1}(0)\cdot \mathbf{p}_{j}^{\prime}(t)\rangle =\langle \mathbf{p}_{1}(0)\cdot \mathbf{p}_{1}(0)\rangle $. This is because first $\langle \mathbf{p}_{1}(0)\cdot \mathbf{p}_{j}^{\prime \prime }(t)\rangle =0$ as $\mathbf{p}_{j}^{\prime \prime }(t)$ is independent of $\mathbf{p}_{1}(0)$ and second $\sum_{j=1}^{N}\mathbf{p}_{j}^{\prime }(t)=\mathbf{p}_{1}(0)$ as the momentum $\mathbf{p}_{1}(0)$ is
conserved in the system. As a result $C(t)$ is in fact a time-independent constant, suggesting that though the initial moving direction will be forgotten quickly by the tagged molecule itself, it will be well remembered by the whole system in future.

This fact implies that the energy transfer process studied here cannot be a Markov process. To verify that the memory is kept during the energy transferring process, we reproduce Fig. 1(e) in an alternative way: In preparing a gas system in our ensemble as described in Step (ii), we rotate additionally the coordinate system so that the initial moving direction of the tagged molecule is along axis $y$. This is equivalent to considering a subset of Helfand's subensemble where the momentum direction of the tagged molecule is specified as well. Fig. 3 (b) shows the result, from which we can see that $\rho_{e}(\mathbf{r},t)$ is obviously anisotropic and suggests clearly the initial moving direction of the tagged molecule.

\begin{figure}[tbp]
\vspace{0.2cm} \epsfig{file=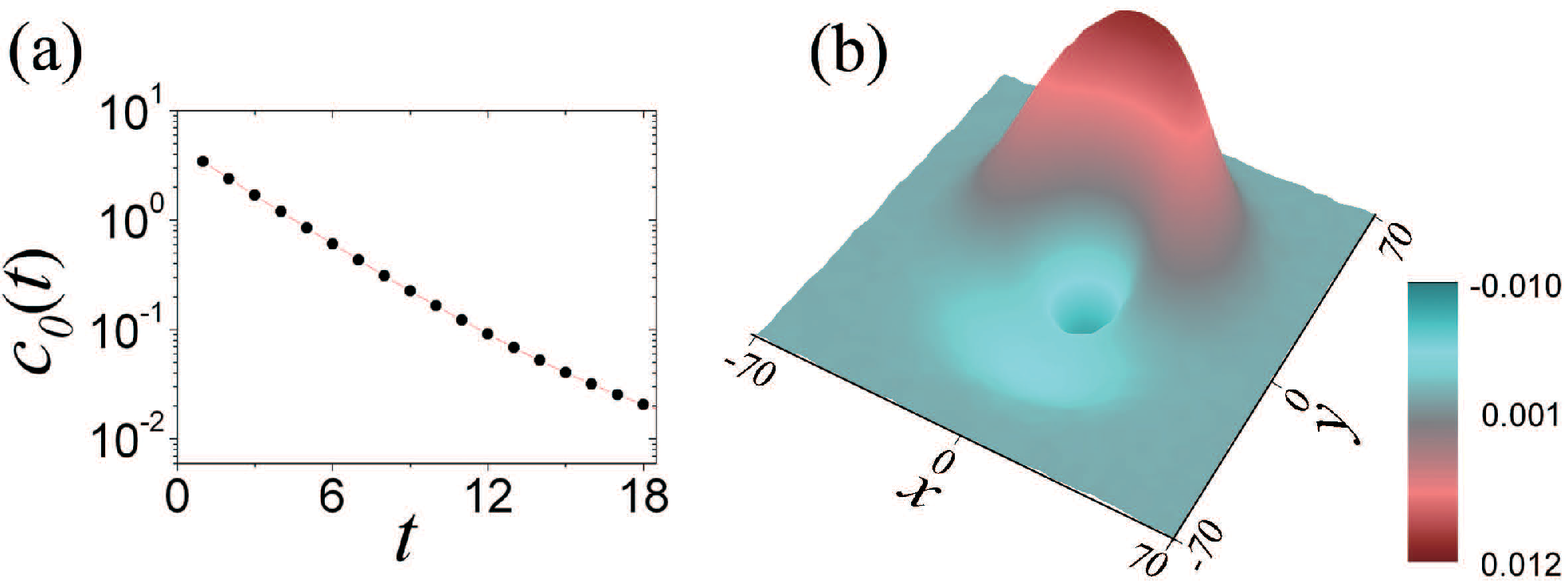,width=8cm}
\caption{(a) The autocorrelation function $A(t)$ of the momentum of the tagged molecule decreases exponentially with time; (b) The same as Fig. 1(e), but the direction of the initial velocity of the tagged molecule is set to be along axis $y$.}
\label{fig3}
\end{figure}

In summary, rather than the Brownian motion, our simulation study of a 2D gas at room temperature shows that the energy carried by a molecule would propagate away in a ballistic wavelike manner. Considering the ensemble average, the profile of the transferred energy is found to be characterized by a ring ridge and a dip in center, and both expand outwards with constant speeds comparable to the speed of sound. The ballistic propagation of the energy is also confirmed by the spatiotemporal correlation function of the energy fluctuation.

We emphasize that our study investigates the energy dispersion of a single molecule in equilibrium state, hence in nature the observed ballistic  characteristic is distinct from the nonequilibrium macroscopic waves such as the sound wave and the heat wave \cite{HW}. For example, the heat wave is a macroscopic relaxing phenomenon, it may decay in fluids due to viscosity and heat conduction, but the energy density distribution in the present study does not decay [see Fig. 2(b)]. The heat wave can also exist in lattice systems \cite{HC1}, but again it only survives for a finite time due to decaying effect. The properties of the energy dispersion studied here are also in clear contrast to those of the macroscopic heat conduction sustained by the temperature gradient; e.g., we have also studied the 3D counterpart of our gas model, and both a 2D and a 3D hard-disc gas model, but obtained qualitatively the same results. However, the heat conduction may have a dramatic dependence on the dimensionality in momentum-conserved systems \cite{HC2}. In spite of these difference, the energy dispersion properties of a molecule must have underlying implications to various macroscopic energy transport behavior. In this respect the mode coupling theory of hydrodynamics, which has been shown powerful in bridging the microscopic and macroscopic descriptions of fluids \cite{hansen}, may provide deep insights.

Finally we would like to suggest a possible laboratory testing scheme of the energy transfer process bases on the energy correlation function discussed (see Fig. 1 (g)-(i) for example simulation results). The requisite laboratory technique is the measurement of the simultaneous position and velocity of a particle immersed in a gas (fluid) \cite{note}. Given this, a sample of $\Delta H_{1}(0)\cdot \Delta H_{j}(t)$ can be obtained by making the measurement to a certain immersed particle at a time and to another after time $t$. Repeating this data collecting process till a sufficient large amount of samples are available; $c_{e}(\mathbf{r},t)$ can then be evaluated.

This work is supported by the National Natural Science Foundation of China under Grants No. 10775115, No. 10975115, and No. 10925525; and the National Basic Research Program of China (973 Program) under Grant No 2007CB814800.


\begin{thebibliography}{99}
\bibitem{note} This required technique has been advanced in a resent laboratory study \cite{scieceexpress}.
\end{thebibliography}
\end{document}